\documentclass[useAMS,usenatbib]{mn2e_alt}
\usepackage{times}
\usepackage{epsfig}

\def\apj{{ApJ}}

\def\apjl{{ApJ}}

\def\aj{{AJ}}
\def\mnras{{MNRAS}}

\def\beq{\begin{equation}}
\def\eeq{\end{equation}}
\def\lapprox{\;\raise.4ex\hbox{$<$}\kern-0.8em \lower .6ex\hbox{$\sim$}\;}

\title[Quasar H~II Regions]
{Quasar H~II Regions During Cosmic Reionization}
\author[Alvarez \& Abel]{Marcelo A. Alvarez\thanks{e--mail:
    malvarez@slac.stanford.edu} and Tom Abel \\
Kavli Institute for Particle Astrophysics and Cosmology, Stanford
University, Stanford, CA 94305}
\date{Accepted 2007 May 31. Received 2007 May 14; in original form
  2007 March 28}
\pagerange{L30--L34}
\pubyear{2007} \volume{380} 

\begin{document}

\maketitle

\label{firstpage}

\begin{abstract}

Cosmic reionization progresses as H II regions form around sources of
ionizing radiation. Their average size grows continuously until they
percolate and complete reionization. We demonstrate how this typical
growth can be calculated around the largest, biased sources of UV emission
such as quasars by further developing an analytical model based on the
excursion set formalism. This approach allows us to calculate the sizes
and growth of the HII regions created by the progenitors of any dark
matter halo of given mass and redshift with a minimum of free parameters.
Statistical variations in the size of these pre-existing HII 
regions are an additional source of uncertainty in the determination of 
very high redshift quasar properties from their observed HII region sizes.
We use this model to demonstrate that the transmission gaps seen in very
high redshift quasars can be understood from the radiation of only their
progenitors and associated clustered small galaxies. The fit sets a lower
limit on the redshift of overlap of $z=5.8 \pm 0.1$. This
interpretation makes the transmission gaps independent of the age of
the quasars observed.  If this interpretation were correct it would
raise the prospects of using  radio interferometers currently under
construction to detect the epoch of reionization.

\end{abstract}


\section{Introduction}
Recent observations are just beginning to reveal the epoch of 
cosmological reionization, which defines a fundamental transition in
the universe, separating the cosmic dark ages (e.g. Rees 1997) from
the epoch of galaxy formation and evolution.  The appearance of a
Gunn-Peterson trough (Gunn \& Peterson 1965) in quasar spectra
indicates that reionization was ending at $z\sim 6$
(e.g. Becker et al. 2001; Fan et al. 2002; White et al. 2003), while
the large-angle 
polarization anisotropy of the cosmic microwave background observed by
the Wilkinson Microwave 
Anisotropy Probe (Spergel et al. 2006) indicates the universe may have been
significantly reionized by $z\sim 10$ (Page et al. 2006). 
Observations of Lyman-$\alpha$ emitting
galaxies at $z\sim 5-7$ are posing puzzles with regard to
the reionization history at those redshifts
(e.g. Haiman 2002; Hu et al. 2002; Malhotra \& Rhoads 2004).

In order to answer these questions much theoretical
effort is underway.  Numerical studies have
provided insights into the asymmetric nature of ionization fronts
(Abel, Norman \& Madau 1999; Ciardi et al. 2001; Alvarez, Bromm, \&
Shapiro 2006a), radiative feedback (Ricotti, Gnedin, \& Shull 2002; 
Shapiro, Iliev, \& Raga 2004; Whalen, Abel, \& Norman 2004; Kitayama
et~al. 2004; Abel, Wise, \& Bryan 2007; Susa \& Umemura 2006; Johnson,
Greif, and Bromm 2006; Ahn \& Shapiro 2007), the reionization history
(Gnedin \& Ostriker 1999; Ciardi, Ferrara, \& White 2003; Sokasian et
al. 2004), and the large-scale structure of reionization (Kohler, Gnedin, \&
Hamilton 2005a; Iliev et al. 2006a; Zahn et al. 2007).  Because
of practical limitations, numerical  studies are expensive
and it is difficult to know which processes to simulate
directly and which to parameterize. Analytical studies can play a
complementary role. Early studies
modelled reionization by considering the
growth of H~II regions around sources of ionizing radiation
in a homogeneous expanding universe with a clumping factor
(e.g. Shapiro \& Giroux 1987).  
While simplistic, models based on this  assumption have proved 
valuable (e.g. Haiman \& Loeb 1997; Madau, Haardt, \& Rees 1999;
Haiman \& Holder 2003; Iliev, Scannapieco, \& Shapiro 2005). 
Studies that describe the thermodynamics of the IGM 
have added additional affects such as non-equilibrium
ionization and heating and an evolving UV background (e.g. Arons \&
Wingert 1972; Shapiro, Giroux, \& Babul 1994; Miralda-Escude \& Rees
1994; Hui \& Gnedin 1997; Miralda-Escude, Haehnelt, \& Rees 2000; Hui
\& Haiman 2003; Choudhury \& Ferrara 2005). 
 
In order to understand the large scale structure
of reionization, analytical models for the sizes H~II regions during
reionization have recently been developed (Wyithe \& Loeb 2004b;
Furlanetto, Zaldarriaga, \& Hernquist 2004 -- hereafter FZH04; Furlanetto
\& Oh 2005; Furlanetto, McQuinn, \& Hernquist 2006a;  Wyithe \& Loeb 2006;
Cohn \& Chang 2007).  FZH04 found that typical H~II region
sizes during reionization were 1-10~Mpc.  These predictions were
qualitatively verified by the radiative transfer simulations of Zahn et
al. (2007).  While they did not directly compare their simulations to the
analytical predictions, they did develop a ``hybrid'' technique and showed
that it gives strikingly similar results to their radiative transfer
simulations.  The FZH04 model has been used to predict the 21--cm
background (McQuinn et al. 2006) as well as the kinetic Sunyaev Zel'dovich
effect (McQuinn et al. 2005).  Furlanetto, Zaldarriaga, \& Hernquist
(2006b) used it to predict the effect of reionization on
Ly$\alpha$ galaxy surveys, while Dijkstra, Wyithe, \& Haiman (2007) used
it to provide a lower limit to the ionized fraction at $z=6.5$. Kramer,
Haiman, \& Oh~(2006) extended the FZH04 model to include the effects of
feedback on the size distribution.
Alvarez et al.~(2006b) used the model to
estimate the cross-correlation between the cosmic microwave and 21--cm
backgrounds on large scales.

Measurement of individual H~II region sizes would provide strong
constraints on the sources 
responsible for reionization.  Until now, such
detection has remained elusive, due to complexities in the
interpretation of quasar spectra.  These difficulties arise
in the interpretation of the ``transmission gap'' between the quasar's
Ly$\alpha$ line and the onset of the Gunn-Peterson trough.  This gap has been 
interpreted as corresponding to the quasar's own H~II region
(e.g., Mesinger \& Haiman 2004; Wyithe \& Loeb 2004a).
Alternatively the gap may be determined by
a combination of the flux from the quasar and the background UV
radiation field -- the Gunn-Peterson trough sets in whenever their
combined flux cannot keep the IGM sufficiently ionized
(e.g., Yu \& Lu 2005; Fan et al.~2006; Bolton \& Haehnelt 2007).  In
this case, the H~II region could be much larger than the size
corresponding to the transmission gap, but note sizes near overlap,
paying particular attention to the effect of recombinations on the mean
free path of ionizing photons, while Lidz, Oh, \& Furlanetto (2006),
Bolton \& Haehnelt (2007), Mesinger \& Haiman (2007), and Maselli et~al.~(2007)
examined the effects on quasar spectra due to density fluctuations in the IGM.

Here we will model the size of
pre-existing HII regions that existed around high redshfit
quasars before they began to shine.
We will use the
conditional H~II region size distribution, which describes
statistically the size distribution of H~II regions that surround
haloes of a given mass.  In \S2 we review the model and derive
the distribution. In \S3 we discuss the
implications for quasar H~II regions, and conclude with a 
discussion in \S4. We adopt parameters based on WMAP 3-year
observations (Spergel et al. 2006), $(\Omega_mh^2$, $\Omega_bh^2$,
$h$, $n_s$, $\sigma_8$)=(0.13,0.022,0.73,0.95,0.74). 

\section{Conditional H~II Region Sizes}
In the model of FZH04, the size of the H~II region in which a point
lies is determined by finding the largest spherical region centered on
the point for which the mean collapsed fraction $f_{\rm coll}>\zeta^{-1}$,
where $\zeta$ is an efficiency parameter which describes how many
ionizing photons are produced per collapsed atom.  By using the
extended Press-Schechter formalism, this condition
can be expressed in terms of the mean overdensity of the region,
$\delta_m$, by
\beq
\delta_m\geq \delta_x(m,z)\equiv \delta_c(z)-\sqrt{2}K(\zeta)\left[
\sigma^2_{\rm min}-\sigma^2(m)\right]^{1/2},
\eeq
where $m$ is the mass of the region, $\sigma^2(m)$ is the variance of
density fluctuations, $\delta_c(z)$
is the threshold overdensity for collapse, and $\sigma^2_{\rm min}\equiv
\sigma^2(m_{\rm min})$ is the variance on the scale of the minimum
halo mass which contributes to reionization, $m_{\rm min}$.
By finding a linear
approximation to the ``barrier'', 
\beq
\delta_x(m,z)\simeq B(m,z)\equiv B_0(z)+B_1(z)\sigma^2(m),
\eeq
where 
\beq B_0(z)=\delta_c-\sqrt{2}K(\zeta)\sigma_{\rm min},
\hspace{1.0cm}
B_1(z)=\frac{K(\zeta)}{\sqrt{2}\sigma_{\rm min}},
\eeq
and $K(\zeta)={\rm erf}^{-1}(1-\zeta^{-1})$,
they were able to derive the mass function of ionized regions,
\beq
M_b\frac{dn}{dM_b}=\sqrt{\frac{2}{\pi}}\frac{\rho_0}{M_b}\left|\frac{d\ln\sigma}{d\ln
    M_b}\right| \frac{B_0}{\sigma(M_b)}{\rm exp}\left[-\frac{B^2(M_b,z)}{2\sigma^2(M_b)}
\right].
\label{mdist_uncond}
\eeq
Additional details can be found in FZH04.

\begin{figure}
\begin{center}
  \includegraphics[width=2.7in]{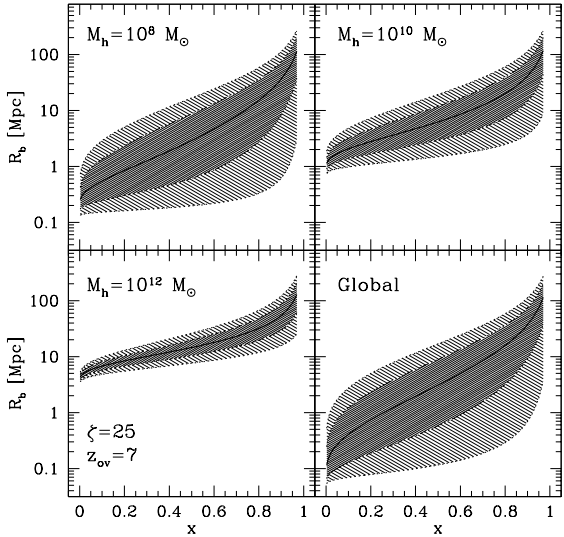}
\caption{Shown is the evolution of the median bubble size (solid),
  along with the 68\% (dark shaded region) and 95\% (light shaded
  region) contours of the distribution.  The ``Global'' panel is for
  the unconditional mean distribution, while the other panels are the
  conditional distributions given that the H~II regions surround halos
  of total mass $M_{\rm h}$, as labelled.  We assumed a minimum source
  halo virial temperature of $10^4$~K. 
\label{halopanels}
\vspace{-0.3cm}
}
\end{center}
\end{figure}

What is the probability,
$f(M_b|M)dM_b$, that a halo of mass $M$ will be located in an ionized
bubble with a size between $M_b$ and $M_b+dM_b$?  This can be
found by considering the halo to be the locus of the first
upcrossing of a random walk in $\delta$. 
First, we determine the conditional probability,
$f(S,\delta_c|S_b,\delta_b)dS$, that a point which first crossed the
barrier at $S_b\equiv \sigma^2(M_b)$ and $\delta_b\equiv
B(M_b,z)$, crosses the halo barrier $\delta_c$ between $S\equiv
\sigma^2(M_h)$ and $S+dS$.  As has been discussed previously
(e.g. Sheth 1998), this can be
accomplished by using the standard expression for the distribution of
up-crossings of a linear barrier for trajectories starting from $S=0$ and $\delta=0$,
\beq
f(S,\delta_c|0,0)dS=\frac{\delta_c}{\sqrt{2\pi S}}{\rm exp}\left[
-\frac{\delta_c^2}{2S}\right]\frac{dS}{S},
\eeq
but moving the origin to $(S_b,\delta_b)$ (see also Furlanetto,
McQuinn, \& Hernquist 2006a, equation 2),
\beq
f(S,\delta_c|S_b,\delta_b)dS=\frac{\delta_c-\delta_b}{\sqrt{2\pi}
  (S-S_b)^{3/2}} 
{\rm exp}\left[-\frac{(\delta_c-\delta_b)^2}{2(S-S_b)}\right]dS.
\eeq
The conditional probability that the first up-crossing of the bubble
barrier occured between $S_b$ and $S_b+dS_b$, given that it crosses
the halo barrier $\delta_c$ at $S>S_b$, is 
\beq
f(S_b,\delta_b|S,\delta_c)dS_b=\frac{f(S_b,\delta_b|0,0)}{f(S,\delta_c|0,0)}f(S,\delta_c|S_b,\delta_b)dS_b.
\eeq
Thus, the probability that a halo of mass $M$ is inside an ionized
bubble of mass between $M_b$ and $M_b+dM_b$ is
\begin{eqnarray}
&&f(M_b|M)dM_b=f(S_b,\delta_b|S,\delta_c)\frac{dS_b}{dM_b}dM_b\\
\nonumber
&=&\frac{1}{\sqrt{2\pi}}\frac{B_0}{\delta_c}
\left[\frac{S}{S_b(S-S_b)}\right]^{3/2}\left|\frac{dS_b}{dM_b}\right|\times\\
\nonumber
&& {\rm
  exp}\left\lbrace\frac{\delta_c^2}{2S}-\frac{\delta_b^2}{2S_b}-\frac{[\delta_c-\delta_b]^2}{2(S-S_b)}\right\rbrace dM_b.
\label{mdist}
\end{eqnarray}
Shown in Fig. \ref{halopanels} are the median, 68\%, and 95\%
contours of the distribution $f(M_{\rm b}|M_{\rm h})$, for different
halo masses $M_{\rm h}$, as well as the ``global'' relationship given
by equation (\ref{mdist_uncond}). The figure shows comoving radius rather than mass,
defined by the relation $R\equiv [3M/(4\pi\rho_0)]^{1/3}$. 

\section{Quasar H~II Regions}

In this section we will discuss the implications of the conditional
H~II region 
size distribution on observations of quasar H~II regions.  Before
proceeding, we briefly describe what we mean by quasar H~II
region, especially near the end of reionization.

Late in reionization, close to percolation, the H~II region sizes grow
rapidly.  This is of course the expected behaviour, but it is not
clear what meaning to attach to any given H~II region when most of the
universe is already ionized.  For H~II regions around very rare quasar
host halos, it is reasonable to expect that the central
H~II region in which the quasar forms is much larger than the
surrounding H~II regions.  While the mean global ionized fraction may
be high, the fluctuations in ionized fraction outside of the central
H~II region are likely to be on much smaller scales.  

\subsection{Interpretation of observed H~II region sizes}
The variation in the size of {\em pre-existing} H~II regions that
surround  
quasars when they begin to shine manifests itself in a theoretical 
uncertainty in the determination of properties of quasars, such as their
ionizing photon luminosity and age, as well as properties of the surrounding
medium, such as the neutral fraction and gas density.  For a quasar
that has been on for a  
time $t$ and has an ionizing photon luminosity $\dot{N}_\gamma$, the
observed line-of-sight radius of its H~II region, $R_{\rm obs}$, is
related to the radius of the H~II region in which it initially lies,
$R_{\rm b}$, by  
\beq
\frac{\dot{N}_{\gamma}t}{(1-x)n_H}\equiv V=\frac{4\pi}{3}\left(R_{\rm
  obs}^3-R_{\rm b}^3\right)=V_0\left(1-R_{\rm b}^3/R_{\rm obs}^3\right),
\eeq
where $V_0\equiv 4\pi R_{\rm obs}^3/3$ is the volume enclosed by the
observed H~II region,
$n_H$~is the hydrogen number density, $x$ is the ionized
fraction, and we have assumed that the apparent age of the quasar, $t$,
is less than the recombination time, $t_{\rm 
  rec}= 1.9~{\rm Gyr}~C_l^{-1}(1+\delta)^{-1}[(1+z)/7]^{-3}$, where
$C_l$ is the clumping factor.  Because the H~II region size is
measured along the line of sight, this equation is correct even when
the ionization front velocity is relativistic (e.g. White~et~al.~2003;
Shapiro~et~al.~2006).
The quantity $V$ is the volume that is
actually ionized by the quasar itself, {\em excluding} the volume
already ionized by existing nearby sources, $4\pi R_{\rm b}^3/3$.  

\begin{figure}
\begin{center}
  \includegraphics[width=2.7in]{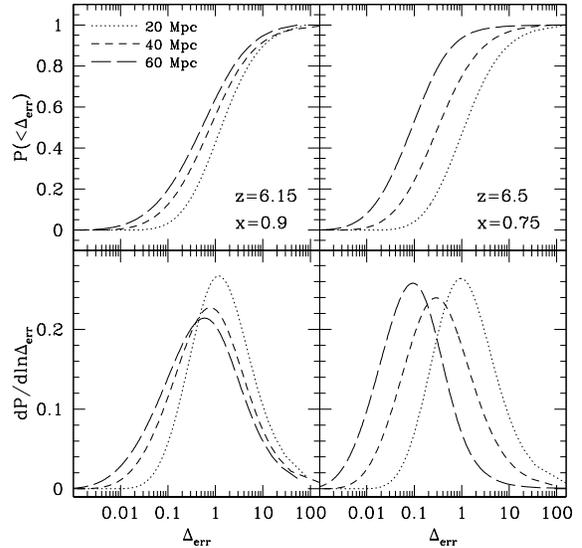}
\caption{Distribution of error in volume ionized by the quasar,
  $\Delta_{\rm err}$ (equation \ref{error}), for a given observed
  comoving H~II region radius, $R_{\rm 
  obs}=20$~Mpc (dotted), 40~Mpc (dashed), and 60~Mpc (long
  dashed), for a central halo mass $M_{\rm h}=10^{12} M_\odot$, a
  minimum source halo virial temperature of $10^4$~K, and and
  efficiency $\zeta=15$.  Thus upper
  panels are the cumulative distribution, while the lower panels are
  the differential distribution.
\label{robs}
\vspace{-0.3cm}
}
\end{center}
\end{figure}
What is the uncertainty in the determination of $V$,
given that the initial radius of the H~II region, $R_{\rm b}$, is
unknown?  We express this uncertainty in terms of an error
\beq
\Delta_{\rm err}\equiv \frac{V_0-V}{V},
\label{error}
\eeq
which is the fractional amount by which the uniform-IGM estimate for the
volume ionized by the quasar, $V_0$, exceeds the actual one, $V$. We
obtain 
\beq
\frac{dP}{d{\rm ln}\Delta_{\rm err}}=\frac{1-R_{\rm b}^3/R_{\rm
    obs}^3}{3(2-R_{\rm b}^3/R_{\rm obs}^3)}\frac{dP}{d{\rm ln}R_{\rm b}},
\eeq
where $P$ is the cumulative probability, e.g. $P(<M_{\rm
  b}|M_{\rm h})=P(<R_{\rm b}|M_{\rm h})=\int_0^{M_{\rm b}}f(M|M_{\rm
  h})dM$, 
$f(M|M_{\rm h})$ is the conditional bubble size distribution as
defined in equation (8), 
$M_{\rm b}\equiv 4\pi\rho_0 R_{\rm b}^3/3$, and $M_{\rm
  h}=10^{12} M_\odot$.  
The error distribution is shown in Fig. \ref{robs} for different values
of $R_{\rm obs}$.  For larger observed H~II regions, the overestimate
of $V$ is smaller.  If the ionizing luminosity and age of the quasar are known,
as well as the density of the surrounding gas, then the neutral 
fraction can be determined according to 
\beq
1-x\equiv x_{\rm HI}=\frac{\dot{N}_\gamma t}{n_HV}.
\eeq
Of course, the luminosity, age, and density are not known exactly, but
arguments can be made based upon their likely values (e.g.~Wyithe \&
Loeb 2004a).  The error in $V$ is related to an error in
$x_{\rm HI}$ by $x_{\rm HI}/x_{\rm HI,0}=1+\Delta_{\rm err}$, where
$x_{\rm HI,0}$ is the neutral fraction inferred by assuming the quasar
began to shine in a completely neutral medium.  Since
$\Delta_{\rm err}>0$, the neutral fraction is always underestimated
when the presence of existing H~II regions is not taken into account.
There are therefore two reasons why higher neutral fractions lead to
smaller H~II regions.  First, higher neutral fractions lead to slower
ionization fronts propagating away from the quasar and thus to smaller
H~II regions.  Second, the higher the neutral fraction, the smaller
the pre-existing H~II region, which also leads to a smaller observed
H~II region. 

\subsection{Proximity zones in quasar spectra}
\begin{figure}
\begin{center}
  \includegraphics[width=2.8in]{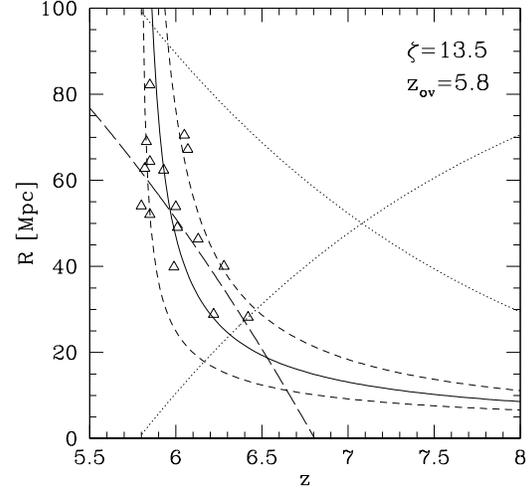}
\caption{Median (solid) and 68\% contours (dashed) of the size
  of H~II regions surrounding a $10^{12} M_\odot$ halo vs. redshift
  for a model with $\zeta=13.5$ and $z_{\rm ov}=5.8$. 
  Also shown are the ``proximity zone'' radii (triangles) of
  sixteen quasars as determined by Fan~et~al.~(2006).  
  The dotted lines are the ionized and neutral fractions multiplied by
  100.  Shown also is a linear fit (long dashed) of the proper size vs. redshift, as in Fan~et~al.~(2006).
\label{fan}
\vspace{-0.3cm}
}
\end{center}
\end{figure}

Fan~et~al.~(2006) define ``proximity zones'' as regions
where the transmission is greater than ten percent, but do not
explicitly associate them with the size of the quasar's H~II
region. These proximity zones have radii which increase
steeply over the range $6.4\ga z\ga 5.8$, from $R_{\rm p}\simeq 30$ to
$R_{\rm p}\simeq 80$ comoving Mpc (Fig. \ref{fan}) . These values are 
similar to the size of H~II regions around $10^{12}M_\odot$-halos (a
likely minimum value for host halo masses of observed quasars at 
$z\sim 6$; e.g. 
Volonteri \& Rees 2006; Li et al.~2006) near the end of reionization 
(Fig.~\ref{halopanels}).

Shown also in Fig. \ref{fan} is the typical size of H~II
regions surrounding halos with masses $M_{\rm h}=10^{12}M_\odot$ for a model
in which the ionized fraction reaches unity at $z_{\rm ov}=5.8$ and the 
integrated Thomson scattering optical depth (assuming once ionized helium) 
is $\tau_{\rm es}\simeq 0.055$.  The
theoretical expectation for the evolution H~II region sizes agrees quite 
well with the measured proximity zone sizes.  This suggests that the
evolving sizes of the proximity zones measured by Fan~et~al.~(2006)
can be explained by the growth of cosmic H~II 
regions driven by clustered sources around the quasars alone.
Taking into account the contribution from the
quasars could change the theoretical prediction, since the flux
contributed by the quasar at distances greater than the size of the
pre-existing bubble may be large enough to increase transmission
there. For a quasar with a spectral shape $L_\nu\propto \nu^{-1.5}$ (e.g., 
Bolton \& Haenelt 2007) and ionizing photon luminosity $\dot{N}$, 
ionization equilibrium implies that the neutral fraction at a comoving 
distance $R$ is 
\beq
f_{\rm HI}\approx 4\times 10^{-6}
\left(\frac{c_l}{10}\right)^{}
\left(\frac{R}{38 \ \rm Mpc}\right)^2
\left(\frac{\dot{N}}{2\times 10^{57}{\ \rm s}^{-1}}\right)^{-1},
\eeq 
while the neutral fraction necessary to obtain a Gunn-Peterson optical
depth $\tau_{\rm GP}$ is given by (Fan et al. 2006),
\beq
f_{\rm HI}\approx 4\times 10^{-6} 
\left(\frac{\tau_{GP}}{2}\right)
\left(\frac{1+z}{7}\right)^{-3/2}.
\eeq 
For reasonable IGM clumping factors and quasar ionizing photon
luminosities, therefore, the quasar by itself is able to cause
transmission at a distance of 40 comoving Mpc, very similar to the size of
the transmission gaps of the highest redshift quasars measured by
Fan~et~al.~(2006).  Thus, there are two possible explanations for the
rapid increase in the size of the trasnmission gaps: 1) HII 
regions, defined as regions where hydrogen is mostly ionized, are much 
larger than the observed transmission gaps at $z<6.4$.  The transmission 
gaps correspond to smaller regions, within which the evolving UV intensity 
from a combination of nearby galaxies and the quasar ionizes the IGM to a 
level sufficient to allow transmission, or 2) The transmission gaps 
correspond to the extent of the HII regions themselves, which grow rapidly as 
overlap takes place at $z\sim 6$.  The UV background intensity within 
these HII regions is dominated by galaxies clustered around the central 
quasar and is sufficient to cause the observed transmission.

\section{Discussion}
We have used the conditional H~II region size distribution around
halos with a mass $\approx 10^{12}~M_\odot$ to model observations of quasar
H~II regions.  Our results can be summarized as follows:
\begin{itemize}
\item[1.] Due to the biased location of quasars, the H~II regions that
  surround them just before they begin to emit ionizing radiation are
  likely to be large, with radii of order tens of Mpc, even when
  the mean ionized fraction is of order thirty percent.
\item[2.] For observed quasar H~II regions with sizes of order tens of
  Mpc, it is difficult to determine the properties of the surrounding
  medium, due to uncertainties in the size of 
  pre-existing H~II regions --  neutral fractions will be
  underestimated for a given quasar luminosity and lifetime.  This
  effect is strongest for small observed H~II regions around quasars
  late in the reionization epoch. 
\item[3.] The observed transmission gaps around high-redshift 
quasars may have a direct correspondence to cosmic H~II regions created by
galaxies clustered around the central quasar.  The steep increase in the
transmission gap size at $z<6.4$ can be explained by the rapid growth of
these H~II regions at the epoch of overlap.

\end{itemize}
We expect conclusion 1 to be generally true, but conclusions 2
and 3 require further explanation. 

For conclusion 2, the situation is more complicated when quasar H~II
region sizes are obtained from transmission gaps in their spectra.
In this case, effects such as a boost in the neutral hydrogen
abundance due to clumping and Ly$\alpha$ absorption from the damping
wing of neutral gas outside the H~II region can cause the
inferred H~II region size to be smaller than it actually is (Bolton \&
Haehnelt~2007; Maselli et al.~2007).  This underestimate leads to an
overestimate of neutral hydrogen abundance for 
a given quasar luminosity and lifetime.  Neglecting the H~II region
created by the nearby galaxies and their progenitors, on the other hand,
leads to an underestimate of the neutral fraction.  A natural question
is which of these two competing effects is stronger.  For example,
Maselli et~al.~(2007) find that the physical radius corresponding to the
transmission gap is about 40 percent larger than that of the H~II region,
which corresponds to an underestimate of its volume by a factor of
3, and an overestimate of the neutral fraction by that
same factor.  In order to offset this effect, $\Delta_{\rm err}>2$
would be necessary.  For example, the probability that a quasar
surrounded by a 40 comoving Mpc H~II region will have $\Delta_{\rm err}>2$ is
about 30 percent, for $x=0.9$ at $z=6.15$ with $T_{\rm min}=10^4$~K, as
shown in Fig. 2. At earlier times and lower
ionization fraction, $x=0.75$ and $z=6.5$, the probability is
only ten percent.  The two competing effects are thus similar in
magnitude when the ionized fraction is high, $x\sim 0.9$, but the
``apparent shrinking'' reported by Maselli et al. (2007) is likely to
dominate at earlier times, when $x\lapprox 0.5$.
These estimates are quite uncertain, however  -- rarer
sources and/or quasar hosts could make these probabilities higher by
increasing the pre-existing bubble size, and more detailed
calculations will be necessary to refine these estimates
further.   

For conclusion 3 to be valid, it is necessary for the UV radiation field within
ionized bubbles to keep the volume-averaged neutral
fraction low enough to create the observed transmission gaps.
This is plausible, since the UV radiation field near
quasars is likely enhanced due to their biased environment (e.g., Yu
\& Lu 2005),
with a size approaching that of the mean free path imposed by
Lyman-limit systems ($\sim~50$ comoving Mpc at $z\sim6$; Gnedin \& Fan
2006).

There have been recent suggestions that detection of large ($\sim 50$
comoving Mpc) H~II regions by 21--cm tomography can reveal the
locations of quasars and bright galaxies that just underwent an AGN 
phase (Wyithe, Loeb \& Barnes 2005; Kohler et al. 2005b).  Our results here 
suggest that large, H~II regions may be visible at relatively low 
redshifts, $z\sim 7$, where foreground contamination of the 21--cm 
observations is lowest.  Thus, the 21--cm observations 
hold the most promise for descriminating between the two possible 
explantations for the evolution of transmission gap sizes 
discussed here.  In addition, measurement of the sizes of H~II regions 
around bright quasars by 21--cm tomography, when combined with the 
conditional H~II region size distribution, will provide powerful 
constraints on the theory of reionization and the nature of high-redshift 
quasars.

In our model the global ionized
fraction is proportional to the collapsed fraction, so that the same
ionizing efficiency is assigned to collapsed matter, regardless of halo
mass, epoch, or environment.  In such a model, the ionized fraction grows
exponentially with time.  The reionization history is likely to be more
complex than this, especially in light of the optical depth measured
by WMAP.  For example, self-regulated reionization, in which the lowest
mass objects do not produce ionizing photons when they form within
existing H~II regions, can extend the reionization epoch, relieving the
tension between a percolation at $z\sim 6$ and the WMAP value of
$\tau_{\rm es}\simeq 0.09$ (Haiman \& Bryan 2006; Iliev et al.~2006b).  
We note, however, that our best-fitting model to the evolution of the
transmission gap sizes, in which overlap is complete at $z\simeq 5.8$ and
$\tau_{\rm es}\simeq 0.055$, is in marginal agreement with the value
obtained from WMAP 3 year polarization data, $\tau_{\rm
  es}=0.088^{+0.028}_{-0.034}$ (Spergel et al.~2006).  This indicates
that the level of early 
reionization demanded by the WMAP measurement may be quite modest, while
still allowing for a late overlap at $z=5.8$.

The fact that H~II regions are likely to surround quasars before they
begin to shine, and that their size will have a strong
redshift dependence, are likely to be crucial in developing a more
complete understanding of observations of the high redshift universe.

\vspace{-0.5cm}
\section*{Acknowledgments}
We wish to thank Zoltan Haiman and Peng Oh for a careful reading of an 
early draft, and Xiaohui Fan and Miguel Morales for helpful
discussion. This work was partially supported by NSF CAREER award AST-0239709
from the National Science Foundation.

\label{lastpage}

\end{document}